\documentclass[aps,pra,twocolumn,showpacs,preprintnumbers,amsmath,amssymb,footinbib]{revtex4}
\usepackage[utf8]{inputenc}
\usepackage{amsmath}
\usepackage{amssymb}
\usepackage{color}
\usepackage{epsfig}
\usepackage{siunitx}
\usepackage{tikz,graphicx}
\renewcommand{\arraystretch}{2}
\newcommand{\sy}[2][+]{\Gamma^{#1}_{#2}} % print representation names (default with positiveparity)
\newcommand{\cuo}{Cu$_{\mathrm{2}}$O}

\newcommand{\br}{\boldsymbol{r}}
\newcommand{\bo}{\boldsymbol{o}}
\newcommand{\bc}{\boldsymbol{c}}
\newcommand{\bp}{\boldsymbol{p}}

\newcommand{\bR}{\boldsymbol{R}}
\newcommand{\bF}{\boldsymbol{F}}
\newcommand{\brho}{\boldsymbol{\varrho}}

\newcommand{\opH}{\mathcal{H}}
\newcommand{\den}{\rho_{ci}}

\newcommand{\nmax}{n_{\mathrm{max}}}

\begin{document}
\title{Interaction of charged impurities and Rydberg excitons in cuprous oxide}

\author{Sjard Ole \surname{Kr{\"u}ger}}
\email[]{sjard.krueger@uni-rostock.de}
\author{Heinrich \surname{Stolz}}
\author{Stefan \surname{Scheel}}
\affiliation{Institut f{\"u}r Physik, Universit{\"a}t Rostock, 
Albert-Einstein-Stra{\ss}e 23-24, D-18059 Rostock, Germany}

\date{\today}

\begin{abstract}
We investigate the influence of a static, uncorrelated distribution of charged 
impurities on the spectrum of bound excitons in the copper oxide {\cuo}. 
We show that the statistical distribution of Stark shifts and ionisation rates 
leads to the vanishing of Rydberg resonances into an apparent continuum. The 
appearance of additional absorption lines due to the broken rotational symmetry,
together with spatially inhomogeneous Stark shifts, leads to a modification of 
the observed line shapes that agree qualitatively with the changes observed in 
the experiment.
\end{abstract}

\maketitle

\section{Introduction}
Semiconductor Wannier excitons are quasi-particles comprised of an electron and 
a hole bound by their mutual Coulomb interaction \cite{wannier1937}. These 
states have first been observed in the 1950s in {\cuo} 
\cite{gross1952,gross1956}, where they appear as a series of resonances below 
the band gap and show remarkable resemblance to the hydrogenic Rydberg series. 
Recently, excitons with large principal quantum numbers of up to $n=25$,
termed Rydberg excitons, have been observed in {\cuo} \cite{kazimierczuk2014}. 
These Rydberg excitons are very sensitive to perturbations of their 
surroundings, just like their atomic counterparts. For example, an 
intensity-dependent bleaching of the resonances has been observed 
\cite{kazimierczuk2014}, which has been interpreted as an excitonic Rydberg 
blockade. Furthermore, the deviations of the spectrum of Rydberg excitons from 
a purely hydrogenic series can be combined into a (phenomenological) quantum 
defect $\delta_{n,\ell}$ that is induced by the nonparabolic hole dispersion and 
other central-cell corrections 
\cite{schoene2016_prb,schweiner2016impact,alvermann2018}.

Since their first observation, the influence of electric and 
magnetic fields on Rydberg excitons
\cite{schweiner2017magnetoexcitons,kurz2017,heckotter2018_2,zielinski2019, artyukhin2018}, the 
mutual dipole-dipole interaction between them \cite{walther2018}, their 
fluorescence \cite{takahata2018.1} and inter-excitonic transitions 
\cite{krueger2019} have been studied. Additionally, proposals have been put 
forward to use them for the implementation of masers 
\cite{ziemkiewicz2018,ziemkiewicz2019} as well as the realisation of topological 
spin phases in lattice potentials \cite{poddubny2019}.

Another effect that has sparked substantial interest is the perturbation by 
free carriers, i.e. the electron-hole plasma \cite{heckoetter2018,semkat2019}. 
It has been observed that the introduction of an electron-hole plasma by 
pumping above the band gap leads to a Mott transition for the Rydberg excitons.
There, the band gap is lowered but the positions of the excitonic resonances 
are almost unaffected until the band gap crosses them, and the resonances 
vanish into the ionisation continuum \cite{heckoetter2018}. The apparent 
suppression of the highest exciton resonances follows a similar phenomenology 
as the Rydberg blockade mechanism. In addition to this plasma-induced shift of 
the band gap, the experiments revealed a static shift that is already present
without the introduction of free charge carriers.

It has long been proposed that disorder introduced, e.g., by charged impurities 
might lead to the appearance of an exponential decay of the absorption coefficient
below the band gap, as well as a shifted band gap \cite{dow1972}. Charged impurities
may form in a compensated semiconductor, containing both acceptors and donors, as it
can be energetically favourable for donor-acceptor pairs to ionise if their binding
 energies are sufficiently low $E_A + E_D <E_g$ \cite[Sec. 7.1.3]{cardona}.
The main point defects present in a typical {\cuo} crystal are Cu and O vacancies
\cite{biccari, ito1997} fulfilling this condition and acting as acceptors and donors,
respectively. Depending on the density of these defects, however, the oxygen vacancies
may form stable charged compound defects $W^{+}$ with copper vacancies
\cite{biccari, zouaghi1972} deep inside the band gap. In order for charge neutrality to
 be upheld, they would have to be compensated by an excess of charged
copper vacancies $V_{\text{Cu}}^{-}$ or free electrons.
These charged impurities introduce a static electric field in which the exciton resonances may
shift or ionise. This can result in a downward shift of the edge of the absorption
continuum which can be interpreted as a reduction of the band gap. The influence of
the static  charged impurities can, at least for low densities of the impurities, be 
modeled by methods originally derived for atomic systems in ionic plasmas, the 
micro-field distributions. These describe the statistical distribution of local 
electric fields which in turn induce statistically distributed Stark shifts and 
ionisation broadening for the excitonic states. In the absence of a screening 
plasma, the suitable micro-field distribution is the one derived by Holtsmark 
in 1919 \cite{holtsmark1919} as all assumptions (static, uncorrelated 
and (locally) homogeneous charge distribution) should be fulfilled.
We will therefore use it in this work to assess the influence of 
charged impurities on the absorption spectra of the Rydberg excitons
in {\cuo}.

The article is structured as follows: In Sec. \ref{sec:sec-I} we describe 
the modeling Hamiltonian used as well as the Holtsmark micro-field distribution
and discuss the central assumptions. In Sec. \ref{sec:sec-II}, 
we will present the numerical spectra, analyse their line parameters and compare
them to experimental spectra. Finally, we will provide a discussion of our results
and an outlook in Sec. \ref{sec:sec-III}.

%%%%%%%%%%%%%%%%%%%%%%%%%%%%%%%%%%%%%%%%%%%%%%%%%%%%%%%%%%%%%%%%%%%%%%
\section{Theory of Stark-shifted excitons}\label{sec:sec-I}

The real-space Wannier equation for an exciton perturbed by external, static 
charges has the form
\begin{align}
\left[\opH_0+\frac{e^2}{4\pi\varepsilon}\sum\limits_i s_i 
\left\{\frac{1}{|\br_e-\bR_{i}|}-\frac{1}{|\br_h-\bR_{i}|}\right\}\right]
\phi(\br_e,\br_h)\\[2pt]
= E\,\phi(\br_e,\br_h),\nonumber
\end{align}
where $\opH_0$ is the unperturbed excitonic Hamiltonian, $s_{i}= \pm 1$ is the 
sign of the perturbing charge and $\varepsilon = \varepsilon_0\varepsilon_r$ is 
the crystal permittivity with $\varepsilon_r = 7.5$ \cite{carabatos1968}.
Furthermore, $\br_{e/h}$ denote the coordinates of the electron and hole that 
form the exciton, and the $\bR_{i}$ are the coordinates of the charged 
impurities. Focussing on only one of the charges, introducing center-of-mass 
and relative coordinates $\bR$ and $\br$, respectively, as well as $\brho_{i} = 
\bR-\bR_{i}$ gives for the interaction Hamiltonian $\opH_{i}$
\begin{equation} 
\opH_{i} = \frac{e^2}{4\pi\varepsilon} s_i 
\left\{\frac{1}{|\brho_{i}-\alpha\br|}-\frac{1}{|\brho_{i} + \beta\br|}\right\}
\end{equation}
where $\alpha=m_e/(m_e+m_h)$ and $\beta = m_h / (m_e + m_h)$ are the relative 
electron and hole masses, respectively. A Taylor expansion around $\br=0$ up 
to first order in $\br$ yields
\begin{equation}
\opH_{i} \approx \frac{e^2}{4\pi\varepsilon} s_i 
\frac{\brho_{i}\cdot\br}{|\brho_{i}|^3}\label{eq:taylor}
\end{equation}
and thus
\begin{align}
\left[\opH_0+\frac{e^2}{4\pi\varepsilon}\sum\limits_i s_i  
\frac{\brho_{i}\cdot\br}{|\brho_{i}|^3}\right]\phi(\br) 
\label{eq:stark-hamiltonian}\\[10pt]
=   \left[\opH_0+e\bF\cdot\br\right]\phi(\br)= E\,\phi(\br)\nonumber
\end{align}
where 
\begin{equation}
\bF = \sum\limits_i \bF_{i} 
=\frac{e}{4\pi\varepsilon}\,\sum\limits_i\,s_i\,\frac{\brho_{i}}{|\brho_{i}|^3}
\label{eq:linear-field}
\end{equation}
is the total electric field of all charged impurities.
Here, the implicit assumption has been made that the length scale on which 
$\bF$ varies is large compared to the extension of excitonic states. In this 
case, the center-of-mass and relative coordinates can be separated if $\opH_0$ 
also permits such a separation, and the truncation of the Taylor expansion 
after the first non-vanishing term is justified. We have tested this assumption
via a Monte-Carlo ansatz, implying that within the range of interest (defined
by the radius of the largest observable excitons), the median relative deviation
from the linear approximation is $<10\%$.

Under the assumption of a static, uncorrelated and homogeneous distribution of 
perturbing charges, the micro-field distribution can be derived from 
Eq.~(\ref{eq:linear-field}), yielding the Holtsmark distribution 
\cite{holtsmark1919,iakubov, pradhan}
\begin{equation}
P(\xi)\,d\xi = \frac{2}{\pi}\, 
\xi\,d\xi\,\int\limits_0^{\infty}\,dx~x\,e^{-x^{3/2}}\,\sin(\xi\,x)
\end{equation}
of the normalised electric field $\xi = |\bF|/F_0$. The normalisation factor
$F_0$ corresponds closely to the field induced by a single impurity at a distance of
 $R_0=\sqrt[3]{3/(4\pi\den)}$
\begin{equation}
F_0 = \frac{e}{2\varepsilon} \left[\frac{4 \den}{15}\right]^{2/3} = \frac{e}{4 
\pi\varepsilon R_0^2}\,\left(\frac{8\pi}{25}\right)^{1/3}\approx\frac{e}{4 
\pi\varepsilon R_0^2},
\end{equation}
where $\den$ denotes the density of charged impurities. $R_0$ coincides roughly
with the average distance of the nearest-neighbour impurity at any point.
The signs of the perturbing charges $s_i$ are irrelevant, as long as the Taylor 
expansion in Eq.~(\ref{eq:taylor}) is limited to the term linear in $\br$. 
Micro-field distributions for more involved scenarios have been derived
including, e.g., a screening plasma and charge-carrier correlations 
\cite{hooper1966,hooper1968}. 

We will focus on the simplest scenario of unscreened charges interacting with 
hydrogen-like excitons fulfilling the nonparabolic Wannier equation
\begin{equation}
\opH_0\,\phi(\br) = \left[\frac{\bp^2}{2\mu} + \Delta T_h(\bp^2) - 
\frac{e^2}{4\pi\varepsilon\, r}\right]\,\phi(\br) = E\, \phi(\br),
\end{equation}
where $\Delta T_h(\bp^2)$ is the nonparabolic part of the hole dispersion which 
is responsible for the excitonic quantum defects. The approach to solve
this equation based on reformulating it as a Sturmian Coulomb problem 
\cite{szmytkowski2012} has been outlined in Ref.~\cite{schoene2016_prb}. The 
relative absorption coefficients of the Stark spectra
$\alpha_{0}(\omega, \bF)$ are then derived by diagonalisation of the 
Wannier equation (\ref{eq:stark-hamiltonian}) in the basis of the 
eigenstates of $\opH_0$.

If the Hamiltonian of the unperturbed exciton has $O(3)$ symmetry as in our 
model, the quantisation axis can be chosen parallel to $\bF$. In this case, the 
excitonic quantisation axes are distributed statistically, which can be taken 
into account by regarding the exciting light field as unpolarised with 
respect to the quantisation axis, which leads to spectra independent of the 
field direction
$\alpha_0(\omega, \bF) = \alpha_0(\omega, F)= A\,\hbar\omega\, \sum_i L_i(\omega, F)$
with some constant $A$.

The resonance line shapes are modeled by asymmetric Lorentzians 
\cite{toyozawa1964}
\begin{equation}
L_i(\omega,F) =   \frac{f_i}{\pi}\,\frac{\frac{\Gamma_i}{2}+ 2 q_i\, 
g\left(\frac{\omega-\omega_i}{\Gamma_i/2}\right) 
(\omega-\omega_i)}{(\omega-\omega_i)^2 + 
\left(\frac{\Gamma_i}{2}\right)^2}\label{eq:asy-lorentz}
\end{equation}
where $f_i$ is the oscillator strength, $\Gamma_i$ the FWHM linewidth, 
$\omega_i$ the frequency and $q_i$ the asymmetry parameter of the $i$-th 
resonance. For isolated $P$-excitons, the asymmetry can be linked to the
frequency dependence of the phononic scattering \cite{toyozawa1964}. In the
following, however, the $q_i$ have to be interpreted as empirical parameters
used to describe and compare the shapes of the absorption lines in the
experimental and numerical spectra as the apparent change in the asymmetry is induced
by the superposition of multiple lines. In our model, all of these parameters
except for the $q_i$ depend on the electric field $F$. The function $g(x)$ has
been chosen as
\begin{equation}
g(x) = \left\{\begin{array}{c c}1 &\text{ if } \,|x|\, \le\, 4\\ 
e^{-\left(\frac{|x|-4}{4}\right)^2} & \text{ else}\end{array}\right. .
\end{equation}
The corresponding line shape resembles an asymmetric Lorentzian with constant 
asymmetry parameter in the vicinity of the resonance, and a symmetric 
Lorentzian far away from it. This line shape has been chosen as the use of 
asymmetric Lorentzians with constant asymmetry $g(x)=1$ leads to a linearly 
decreasing absorption at the band gap due to the long range decay 
$\propto-(\omega-\omega_i)^{-1}$ of all resonances below it. This behaviour is
not observed in the experiment, where the absorption increases linearly at 
the band gap as predicted by Elliot \cite{elliott1957} and reproduced by
symmetric Lorentzians with their long range decay $\propto (\omega-\omega_i)^{-2}$.
With this choice of $g(x)$, the transition between the asymmetric and symmetric 
Lorentzian happens at about $2\Gamma_i$. The choice of the parameter $4$
is somewhat arbitrary but the resulting spectra are not very sensitive to its
exact choice.

The optical transition matrix elements $o_{n,\ell,m}$  from the crystal vacuum
to the eigenstates of $\opH_0$ are proportional to \cite{elliott1957}
\begin{equation}
o_{n, \ell, m} \propto\left\{\begin{array}{l c} \left.\frac{\partial}{\partial 
r}\,R_{n\ell}(r)\right|_{r=0} & \text{ if } \ell = 1\\
 0 & \text{ else },
\end{array}\right.\label{eq:trans-mat-el}
\end{equation}
for unpolarised light, where $R_{n,\ell}(r)$ denotes the radial part
of the real-space envelope function. The relative oscillator strengths
of the Stark excitons can then be calculated via
\begin{equation}
f_i \propto \left|\bc_i \cdot \bo \right|^2
\end{equation}
where $\bc_i$ is the $i$-th algebraic eigenvector of the Hamiltonian
in Eq.~(\ref{eq:stark-hamiltonian}) and $\bo$ is the vector of the
relative transition matrix elements in Eq.~(\ref{eq:trans-mat-el}),
expressed in the chosen basis. The oscillator strengths of the unperturbed
$P$ excitons scale as $f_n\propto (n^2-1)/n^5$ due to the second-class
nature of the transition from the crystal vacuum to the excitonic state
\cite{elliott1957} (i. e. the transition between the pure Bloch states of valence
and conduction band is dipole forbidden at the zone center due to parity).

\newcommand{\fnatomics}{\footnote{Here and in the following $\{S, P, D, F\}$ refers to states with the orbital quantum numbers $\ell=\{0,1,2,3\}$ respectively, as is common in atomic physics. In general, $\ell$ is only a good quantum number if the Hamiltonian is invariant under the full rotation group $SO(3)$, which cannot be the case in a solid state environment. In the cubic symmetry of {\cuo}, however, a spherically symmetric approximation is sufficiently good \cite{schoene2016_prb} as reflected by our choice of $\opH_0$.}}

\begin{figure}[ht]
\includegraphics[width=\columnwidth]{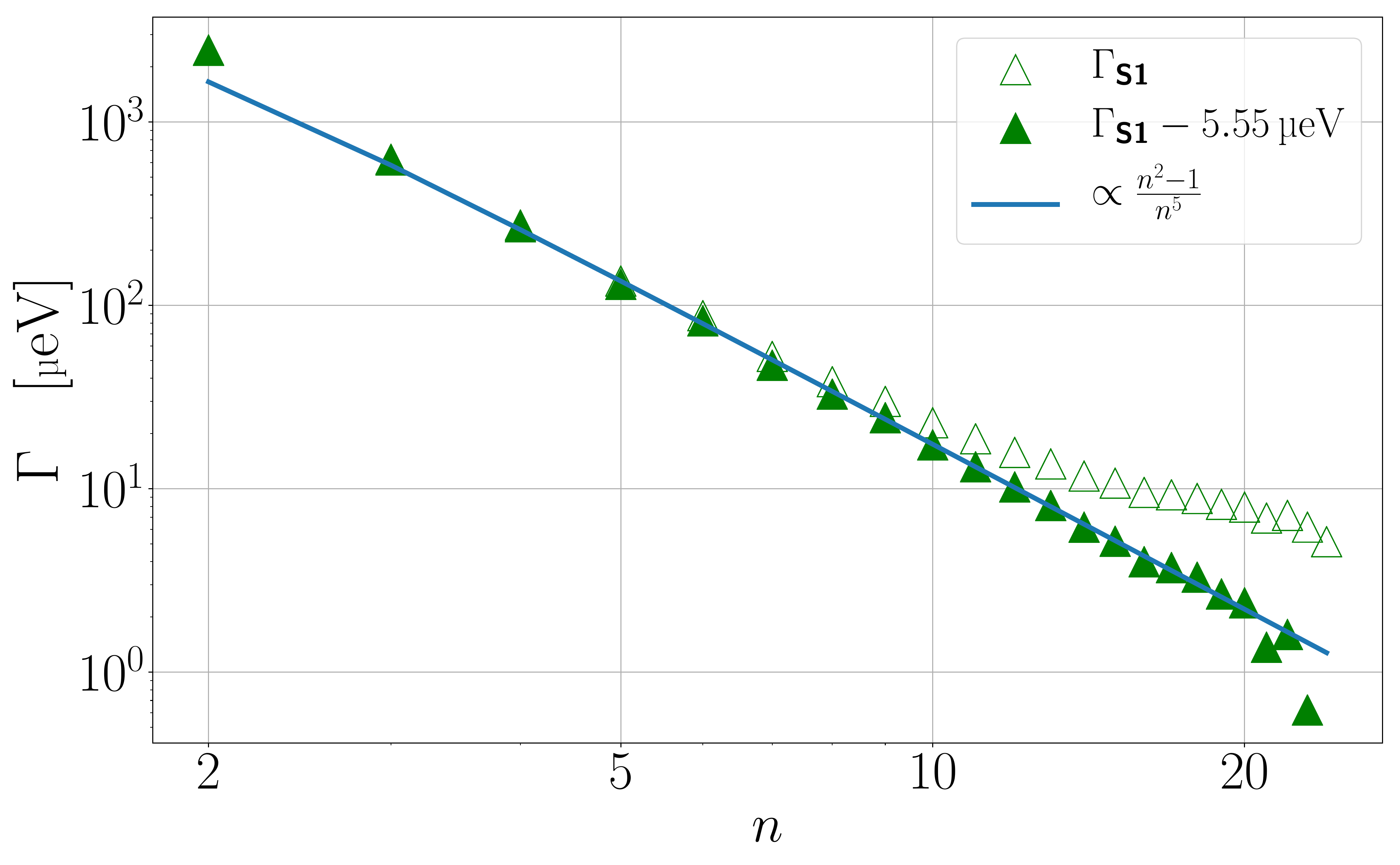}  
\caption{Comparison of the linewidths derived from spectrum \textbf{S1} 
with the theoretically expected scaling (see Sec.~\ref{sec:sec-II}
for details about the experimental spectra). The deviation can be explained
by an additional broadening of $\SI{5.55}{\micro\eV}$ for all lines.}
\label{fig:broadening}
\end{figure}

The linewidths were calculated as laid out in Ref.~\cite{stolz2018} for the 
unperturbed $P$ and $F$ excitons{\fnatomics} (see Tab. \ref{tab:lw}). They
contain all relevant phononic scattering paths into the yellow $1S$ and $2S$
excitons, namely the scattering by LO phonons via the Fr{\"o}hlich mechanism
and the deformation potential scattering by the $\sy[-]{3/5}$ phonons. For the
$S$ and $D$ excitons, the experimental linewidths are not well described by this
theory. We did therefore use extrapolated experimental results from SHG spectra 
\cite{mund2018} for the $3S$ and $3D$ state, respectively, giving
$\Gamma_{nS} = \SI{2}{\milli\eV}\,n^{-3}$ and 
$\Gamma_{nD} = \SI{3}{\milli\eV}\,n^{-3}$. In addition, the complete
experimental spectra seem to be broadened by $5-6\,\si{\micro\eV}$.
Figure~\ref{fig:broadening} shows the experimental linewidths
of one particular absorption spectrum. The deviation from the theoretically 
expected scaling $\propto (n^2-1)/n^5$ already observed in 
Ref.~\cite{kazimierczuk2014} could be explained by the convolution of the 
spectrum with a broadening Lorentzian, whose origin is unknown to us. We modeled 
it by adding $\SI{5.55}{\micro\eV}$ to all input linewidths.

\begin{table}
\def\arraystretch{1.5}
\begin{tabular}{r| c c|| r| c c}
$n$  &$P$ & $F$ & $n$ & $P$  & $F$\\\hline\hline
2 & $\num{1896}$ &   & 3 & $\num{648.4}$ & \\
4 & $\num{268.5}$ & $\num{8.162}$ & 5 & $\num{139.8}$ & $\num{7.489}$\\
6 & $\num{81.68}$ & $\num{5.024}$ & 7 & $\num{51.74}$ & $\num{3.75}$\\
8 & $\num{34.79}$ &  $\num{2.974}$ & 9 & $\num{24.5}$ & $\num{2.369}$\\
10 & $\num{17.89}$ & $\num{1.89}$ & 11 & $\num{13.46}$ & $\num{1.516}$\\
12 & $\num{10.38}$ & $\num{1.226}$ & 13 & $\num{8.17}$ & $\num{1.001}$\\
14 & $\num{6.545}$ & $\num{0.825}$ & 15 & $\num{5.325}$ & $\num{0.687}$\\
16 & $\num{4.389}$ & $\num{0.577}$ & 17 & $\num{3.66}$ & $\num{0.488}$\\
18 & $\num{3.084}$ & $\num{0.416}$ & 19 & $\num{2.623}$ & $\num{0.358}$\\
20 & $\num{2.249}$ & $\num{0.31}$ & 21  & $\num{1.943}$ & $\num{0.27}$\\
22 & $\num{1.691}$ & $\num{0.236}$ & 23 & $\num{1.48}$  & $\num{0.208}$\\
24 & $\num{1.303}$ & $\num{0.184}$ &  & & \\\hline
\end{tabular}
\caption{Input FWHM linewidths in $\si{\micro\eV}$ taken from 
Ref.~\cite{stolz2018}.\label{tab:lw}}
\end{table}

The asymmetry of the lines was taken to be $q_i = -0.24$ for all lines, derived 
from fits to experimental spectra. Clearly, the model Hamiltonian in 
Eq.~(\ref{eq:stark-hamiltonian}) evaluated in a basis of bound excitonic states 
can only be a reasonable description for Stark-excitons that are themselves 
bound. This problem could be addressed by complex scaling techniques 
\cite{zielinski2019} or the introduction of a complex absorbing potential 
\cite{grimmel2017}. Fortunately, the states above the classical ionisation
threshold \cite{gallagher2005}
\begin{equation}
 E_{ion}(F) = -\sqrt{\frac{e^3\,F}{\pi\varepsilon}}
\end{equation}
tend to be broadened by the ionisation as well as the averaging over the micro-field
distribution as they are very sensitive to variations of the electric field.  Our
assumption  is thus, that the ionised states only contribute to a continuous background to 
the absorption spectrum but do not account for prominent absorption lines. Once the Stark
spectra are calculated, the Holtsmark spectra can be derived via
\begin{equation}
\alpha(\omega, \den) = \frac{1}{F_0(\den)}\,\int\limits_0^{\infty} 
dF~P\left(\frac{F}{F_0(\den)}\right) \alpha_0(\omega, F). 
\label{eq:holtsmar-average}
\end{equation}
The integration was performed on a logarithmic grid in $F$ with $F_n/F_{n-1} = 1.001$
for field strengths from $\SI{1}{\milli\volt\per\meter}$ to 
$\SI{100}{\kilo\volt\per\meter}$
via the finite difference summation $\int dF\,f(F) \approx \sum_n\,\Delta F_n\,f(F_n)$
where $\Delta F_n = (F_{n+1}-F_{n-1})/2$.

To summarise this section, the central assumptions of our model are:
\begin{enumerate}
\item The charged impurities are static and their distribution is homogeneous 
and uncorrelated.
\item The electric field induced by the impurities varies on length scales 
considerably larger than the extension of the excitonic states of interest.
\item The spectral structure is dominated by bound excitons below the classical 
ionisation threshold.
\end{enumerate}

%%%%%%%%%%%%%%%%%%%%%%%%%%%%%%%%%%%%%%%%%%%%%%%%%%%%%%%%%%%%%%%%%%%%%%
\section{Numerical results and comparison to experimental 
data}\label{sec:sec-II}

\begin{figure}[ht]
  \begin{minipage}[b]{0.5\textwidth}
	\centering\includegraphics[width=\columnwidth]{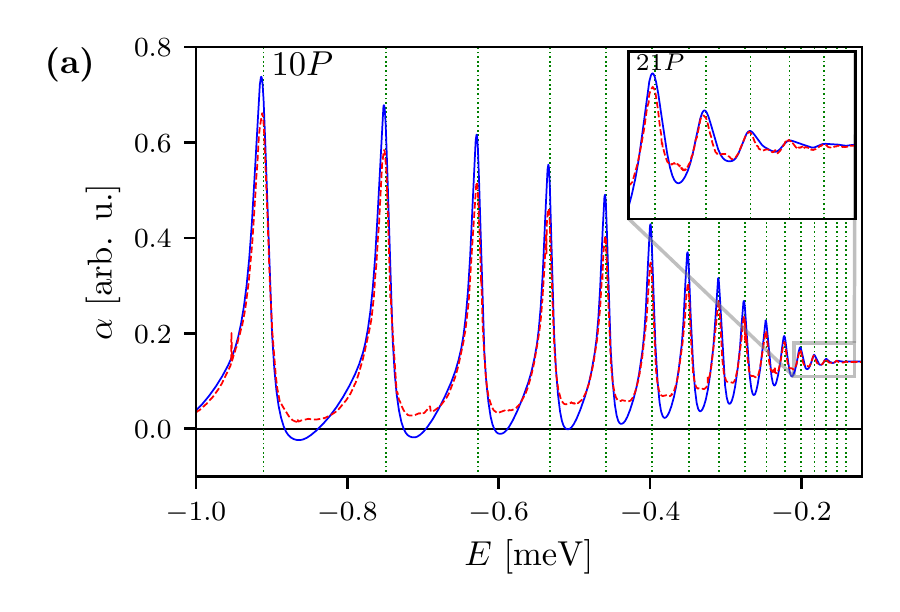}
  \end{minipage}\\
  \begin{minipage}[b]{0.5\textwidth}
	\centering\includegraphics[width=\columnwidth]{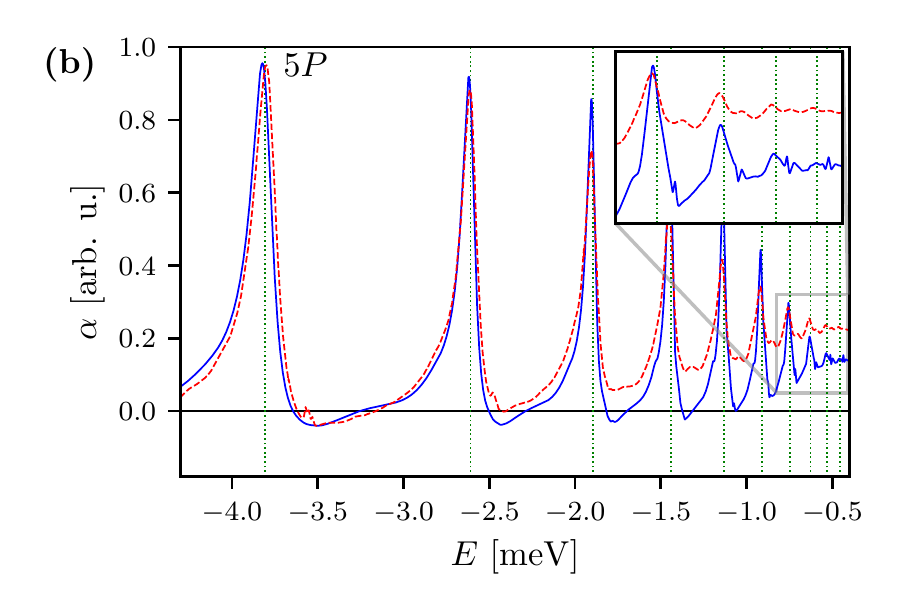}
  \end{minipage}
    
\caption{Comparison of experimental spectra (dashed red lines) and
numerical spectra (solid blue lines). (a): Comparison of spectrum
\textbf{S1} to the numerical spectrum for 
$\den=\SI{1.2e9}{\per\centi\meter\cubed}$.
(b): \textbf{S2} vs. numerical spectrum for $\den = 10^{11}\,\si{\per\centi\meter\cubed}$.
The vertical lines  represent the (numerical) positions of the unperturbed $P$-excitons.}
\label{fig:spectrum}
\end{figure}

We will now apply our numerical method to the resonance spectrum of Rydberg 
excitons and compare with two experimental absorption spectra with
maximum observable principal quantum numbers of $\nmax\approx 25$ (hereafter 
\textbf{S1}) and $\nmax\approx 13$ (hereafter \textbf{S2}). The spectrum 
\textbf{S1} measured at $\SI{1.2}{\kelvin}$ is the one used in the 
Ref.~\cite{kazimierczuk2014} and \textbf{S2} was measured at 
$\SI{1.3}{\kelvin}$. The quantity $\nmax$  denotes the principal quantum number 
above which the resonances form an apparent absorption continuum and no 
individual lines can be resolved. There is, of course, some uncertainty in the 
definition of the highest observable principal quantum number $\nmax(\den)$.
In our analysis, a resonance was considered to have vanished as soon as its
spectral range could not be reliably fitted with the line shape in
Eq.~(\ref{eq:asy-lorentz}). Figure~\ref{fig:spectrum} (a) compares \textbf{S1}
with a numerical spectrum derived for $\den=\SI{1.2e9}{\per\cubic\centi\meter}$
which was chosen to reproduce $\nmax$ while (b) compares \textbf{S2} to a numerical spectrum
for $\den = 10^{11}\,\si{\per\cubic\centi\meter}$. The numerical spectrum in 
Fig.~\ref{fig:spectrum} (b) shows weak additional lines corresponding predominantly 
to the $S$-, $D$- and $F$ excitons, which become dipole allowed due to the 
broken rotational symmetry (see inset). The original experimental spectra
contain a background induced by the phonon-assisted
absorption into the $1S$ and $2S$ states \cite{schoene2017}. This background
has been subtracted for the comparison with the numerical spectra, leading to
the appearance of a negative absorption coefficient on the high-energy side
of the lower resonances.

For the numerical computation, we took into account all states with $\ell \le 
25$, $n_r = n -\ell -1 \le 100$ as well as $m = 0, \pm 1$. This results in basis
sets of dimension $2275$ for $m=0$ and $2175$ for $m=\pm 1$. The calculation can 
be restricted to these three magnetic quantum numbers as the Stark Hamiltonian 
has cylindrical symmetry which ensures that $m$ remains a good quantum number 
(if the quantisation axis is chosen as $z \parallel\bF$) and the optically 
active $P$-states can only be mixed into other states with $m = 0, \pm 1$.

One observes that:
\begin{enumerate}
\item Excitons with large principal quantum numbers smear out and form 
an absorption continuum while the total oscillator strength is conserved.
\item The transition from negative asymmetry parameters $q_n$ for low principal 
quantum numbers to positive ones for the highest $n$, which have been observed
in experimental spectra, is reproduced.
\item Due to the breaking of the rotational symmetries by the Stark effect, 
additional absorption lines -- corresponding to initially dark states -- appear 
in the numerical spectra for high impurity densities.
The strongest additional lines correspond to the $S$-, $D$- and $F$-states.
\end{enumerate}

\begin{figure}[ht]
\includegraphics[width=\columnwidth]{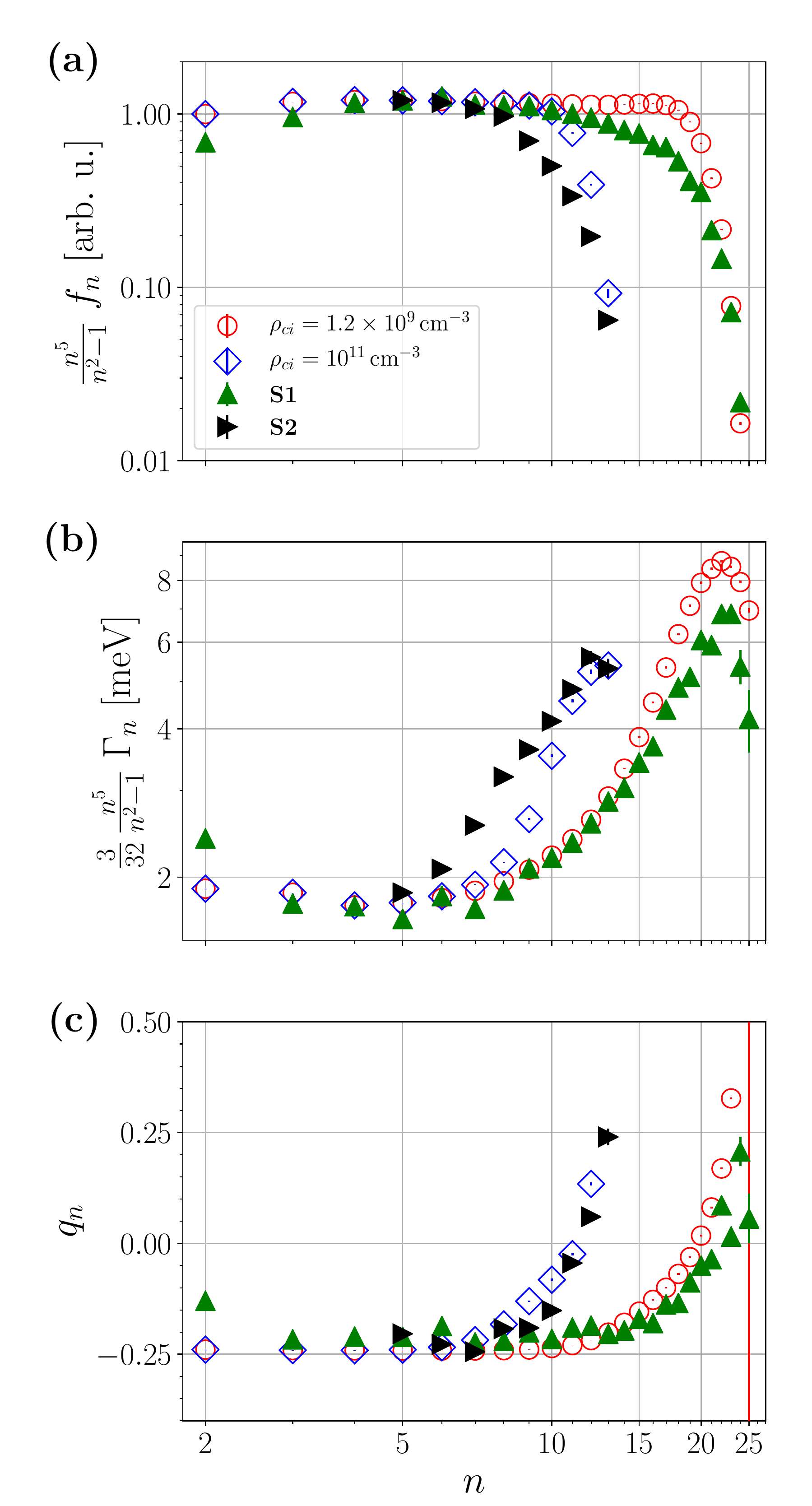} 
\caption{The line parameters of Eq. (\ref{eq:asy-lorentz}) derived by fits to 
the numerical and experimental spectra for different impurity densities: 
  (a) the oscillator strength, (b) the FWHM linewidths and (c) the asymmetry 
parameter.
  The error bars denote one standard deviation.}
\label{fig:line-parameters}
\end{figure}

Figure~\ref{fig:line-parameters} shows the line parameters derived by fits to 
the numerical spectra under the assumption that the underground below each line 
is constant over its width. 
The oscillator strength $f$ [Fig.~\ref{fig:line-parameters} (a)] drops off 
steeply before the lines vanish starting at $n\approx 2\nmax(\den)/3$, an 
observation that could be explained by neither plasma nor phonon interactions 
\cite{semkat2019}. Compared to the experiment, however, the oscillator strength 
follows the $(n^2-1)/n^{5}$ scaling for longer and drops off more steeply for 
large $n$. Note, that the experimental oscillator strengths in 
Fig.~\ref{fig:line-parameters} (a) have been normalised to the numerical ones 
at $n=5$ as their absolute values cannot be compared. 

The FWHM linewidths [Fig.~\ref{fig:line-parameters} (b)] start to deviate from 
the linewidths of the unperturbed resonances $n\approx \nmax(\den)/2$ and drop 
off shortly before $\nmax(\den)$. 
Additional inhomogeneous broadening could be introduced by 
an inhomogeneous straining of the crystal, the ionisation
of states below the classical ionisation threshold \cite{heckotter2018_2}
or the higher orders of the Taylor expansion,
Eq.~(\ref{eq:taylor}), which will become relevant when the electric field 
varies on the length scale of the exciton diameter. 
Furthermore, there 
could be additional sources of micro fields with different micro-field 
distributions, for example optical phonons \cite{dow1972} or surface charges. 
The asymmetry parameter $q$ [Fig.~\ref{fig:line-parameters} (c)] deviates from 
the value for the unperturbed lines for large principal quantum numbers and 
changes sign for $n\approx 4\, \nmax(\den)/5$.

\begin{figure}[ht]
\includegraphics[width=\columnwidth]{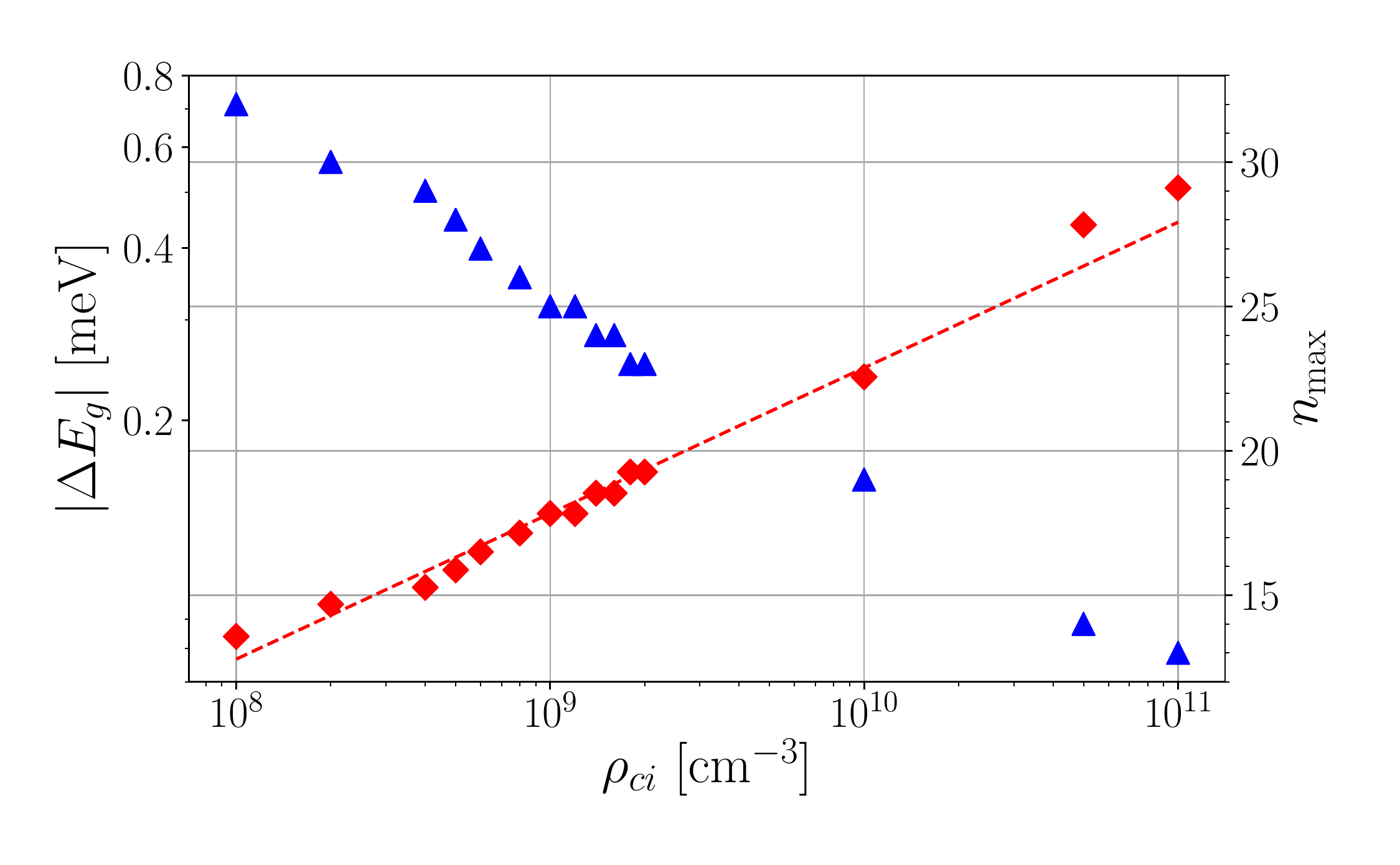}
\caption{The band-gap shift $\Delta E_g(\den)$ (red diamonds) with a power-law 
fit (dashed line) and the maximum observable principal quantum number
$\nmax(\den)$ (blue triangles).}    \label{fig:eg-shift}
\end{figure}

From the numerical spectra, we can extract the maximally observable principal 
quantum number as well as the shift of the band gap.
Figure~\ref{fig:eg-shift} shows the dependence of this band-gap shift on the 
density $\den$ of charged impurities. To a good approximation, it follows a 
power law $\Delta E_g(\den) = -(0.71 \pm 0.17)\,\si{\micro\eV} \, 
(\den/\si{\per\cubic\centi\meter})^{0.254\pm 0.011} 
\propto \den^{1/4}$. This scaling agrees with the dependence of the band-gap shift
on the plasma density derived from many-body theory \cite{semkat2019}. In our case,
however, this is a purely empirical observation. We have not investigated whether this
scaling holds outside the range of impurity densites given here. Clearly, it will
have to break down at some point for large $\rho_{ci}$, as the assumptions made
in the derivation of the Holtsmark distribution break down.

%%%%%%%%%%%%%%%%%%%%%%%%%%%%%%%%%%%%%%%%%%%%%%%%%%%%%%%%%%%%%%%%%%%%%%
\section{Discussion and Outlook}\label{sec:sec-III}

In this work, we have numerically investigated the influence of 
charged impurities on the spectrum of (Rydberg) excitons in {\cuo}. 
Out calculations reproduce experimentally observed phenomena such as the 
vanishing of the resonances with high principal quantum numbers into an 
apparent absorption continuum, accompanied by a drop of the oscillator strength 
of the discernible lines, a broadening as well as a change of the line shape 
towards positive asymmetry parameters $q$.

The breaking of the rotational symmetries inherent in our model leads to the
redistribution of oscillator strength to initially dark states and the corresponding appearance of weak 
additional absorption lines in the spectra for impurity densities greater
$\sim 10^{10}\,\si{\per\centi\meter\cubed}$. In the experimental spectrum 
\textbf{S2} (Fig. \ref{fig:spectrum} (b)) there are indeed some indications
for such peaks, however, the signal/noise ratio of the present spectrum does not allow
for a conclusive analysis. In every case, the non-appearance of such peaks may be
used to establish an upper bound on the charged impurity density of given
crystal samples.

\begin{acknowledgments}
We thank Prof. Manfred Bayer and his group at the TU Dortmund for sharing 
their experimental data and D. Semkat (Greifswald) for helpful discussions.
We acknowledge support by the Deutsche Forschungsgemeinschaft (DFG) within
 the SPP 1929 ``Giant Interactions in  Rydberg Systems (GiRyd)''.
\end{acknowledgments}

\bibliography{sources}
\end{document}